\documentclass[twocolumn,pre,floatfix,superscriptaddress,letterpaper]{revtex4-1}

\usepackage[T1]{fontenc}
\usepackage[latin9]{inputenc}
\usepackage{mathrsfs}
\usepackage{amsmath}
\usepackage{amssymb}
\usepackage{graphicx}
\usepackage{braket}
\usepackage{color}
\usepackage[svgnames]{xcolor}
\usepackage{comment}
\usepackage{natbib}
\usepackage{dsfont}
\usepackage{mdframed}
\usepackage{soul}
\usepackage{pythonhighlight}
\usepackage{placeins}

\makeatletter

\usepackage{listings}

\usepackage{hyperref}

\hypersetup{
  colorlinks=true,
  linkcolor=blue,
  citecolor=ForestGreen,
  urlcolor=DarkOrchid
}

\AtBeginDocument{
  
}

\makeatother

\usepackage{babel}

\newif\ifdraft
\drafttrue
\ifdraft
\newcommand{\note}[1]{ {\textcolor{blue} { **: #1 }}}
\newcommand{\alnote}[1]{ {\textcolor{red} { ***Andre: #1 }}}
\newcommand{\prnote}[1]{ {\textcolor{green} { ***Philipp: #1 }}}
\newcommand{\jknote}[1]{ {\textcolor{blue} { ***Johannes: #1 }}}
\newcommand{\mrnote}[1]{ {\textcolor{ForestGreen} { ***Mauricio: #1 }}}

\else
\newcommand{\note}[1]{}
\newcommand{\alnote}[1]{}
\newcommand{\prnote}[1]{}
\newcommand{\jknote}[1]{}
\newcommand{\mrnote}[1]{}
\newcommand{\lmnote}[1]{}
\newcommand{\jrfnote}[1]{}
\fi
\begin{document}

\title{Optimization of Robot Trajectory Planning with Nature-Inspired and Hybrid Quantum Algorithms}

\author{Martin~J.~A.~Schuetz}
\affiliation{Amazon Quantum Solutions Lab, Seattle, Washington 98170, USA}
\affiliation{AWS Intelligent and Advanced Compute Technologies, 
Professional Services, Seattle, Washington 98170, USA}
\affiliation{AWS Center for Quantum Computing, Pasadena, CA 91125, USA}

\author{J.~Kyle~Brubaker}
\affiliation{AWS Intelligent and Advanced Compute Technologies, Professional Services, Seattle, Washington 98170, USA}

\author{Henry~Montagu}
\affiliation{Amazon Quantum Solutions Lab, Seattle, Washington 98170, USA}
\affiliation{AWS Intelligent and Advanced Compute Technologies,  Professional Services, Seattle, Washington 98170, USA}
\affiliation{AWS Center for Quantum Computing, Pasadena, CA 91125, USA}

\author{Yannick~van~Dijk}
\affiliation{BMW Group, Munich, Germany}

\author{Johannes~Klepsch}
\affiliation{BMW Group, Munich, Germany}

\author{Philipp Ross}
\affiliation{BMW Group, Munich, Germany}

\author{Andre~Luckow}
\affiliation{BMW Group, Munich, Germany}

\author{Mauricio~G.~C.~Resende}
\affiliation{Amazon.com, Inc., Bellevue, Washington 98004, USA}
\affiliation{University of Washington, Seattle, Washington 98195, USA}

\author{Helmut G.~Katzgraber}
\affiliation{Amazon Quantum Solutions Lab, Seattle, Washington 98170, USA}
\affiliation{AWS Intelligent and Advanced Compute Technologies,  Professional Services, Seattle, Washington 98170, USA}
\affiliation{AWS Center for Quantum Computing, Pasadena, CA 91125, USA}
\affiliation{University of Washington, Seattle, Washington 98195, USA}

\date{\today}

\begin{abstract}

We solve robot trajectory planning problems at industry-relevant scales.
Our end-to-end solution integrates highly versatile random-key
algorithms with model stacking and ensemble techniques, as well as path
relinking for solution refinement.  The core optimization module
consists of a biased random-key genetic algorithm.  Through a distinct
separation of problem-independent and problem-dependent modules, we
achieve an efficient problem representation, with a native encoding of
constraints.  We show that generalizations to alternative algorithmic
paradigms such as simulated annealing are straightforward.  We provide
numerical benchmark results for industry-scale data sets.  Our approach
is found to consistently outperform greedy baseline results.  To assess
the capabilities of today's quantum hardware, we complement the
classical approach with results obtained on quantum annealing hardware,
using \texttt{qbsolv} on Amazon Braket.  Finally, we show how the latter
can be integrated into our larger pipeline, providing a quantum-ready
hybrid solution to the problem.

\end{abstract}

\date{\today}

\maketitle

\section{Introduction}
\label{sec:introduction}

The problem of robot motion planning is pervasive across many industry verticals, including (for example) automotive, manufacturing, and logistics. 
Specifically, in the automotive industry robotic path optimization problems can be found across the value chain in body shops, paint shops, assembly and logistics, among others \citep{murad}. 
Typically, hundreds of robots operate in a single plant in body and paint shops alone.
Paradigmatic examples in modern vehicle manufacturing involve so-called welding jobs, application of adhesives, sealing panel overlaps, or applying paint to the car body.
The common goal is to achieve efficient load balancing between the robots, with optimal sequencing of individual robotic tasks within the cycle time of the larger production line.

\begin{figure}[b] 
\includegraphics[width=0.9 \columnwidth]{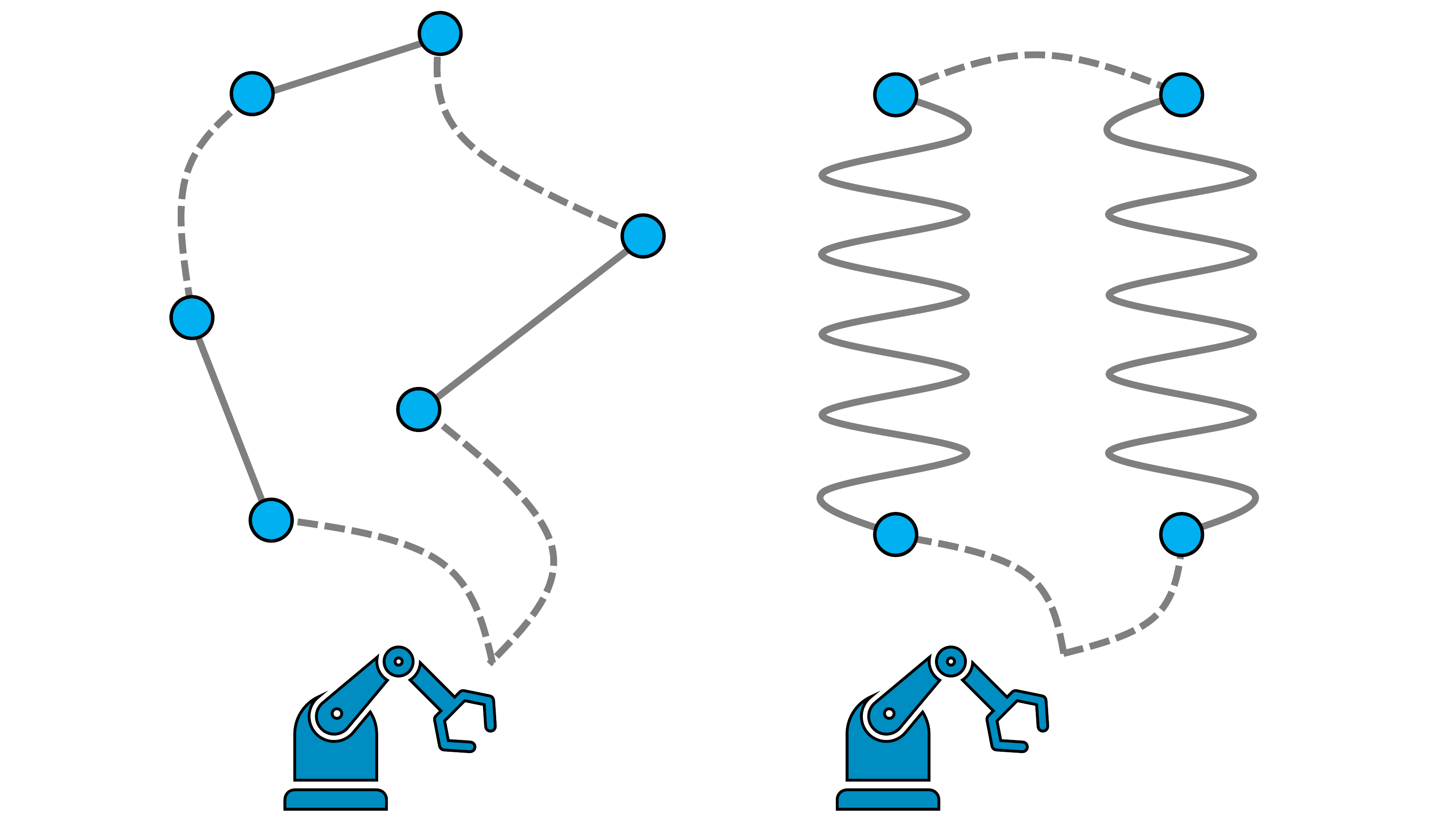}
\caption{(Color online) 
Schematic illustration of the use case.
Robots are programmed to follow certain trajectories along which they apply a PVC sealant along  seams. The seams are highlighted by solid lines with two endpoints each, and are not necessarily straight. Dotted lines represent additional motion between seams.
Every robot is equipped with multiple tools and tool configurations, to be chosen for every seam. 
The goal is to identify collision-free trajectories such that all seams get processed within the minimum time span. 
\label{fig:scheme}}
\end{figure}

Another prototypical example involves a post-welding processes by which every joint is sealed with special compounds to ensure a car body's water-tightness. 
To this end, polyvinyl chloride (PVC) is commonly applied in a fluid state, thereby sealing the area where different metal parts overlap. 
The strips of PVC are referred to as \textit{seams}.
Post application, the PVC is cured in an oven to provide the required mechanical properties during the vehicle's lifetime. 
Important vehicle characteristics such as corrosion protection and soundproofing are enhanced by this process.
Modern plants usually deploy a fleet of robots to apply the PVC sealant, as schematically depicted in Fig.~\ref{fig:scheme}. 
However, the major part of robot programming is typically carried out by hand, either online or offline.
Compared to the famous NP-hard traveling salesman problem, the complexity of identifying optimal robot trajectories is amplified by three major factors.  
First, an industrial robot arm can have multiple configurations that result in the same location and orientation of the end effector.
Furthermore, the PVC is applied with a tool that is equipped with multiple nozzles that allows for application at different angles.
A choice must be made regarding which nozzle to use for seams that display easy reachability.
Finally, industrial robots are frequently mounted on a linear axis; thus, an optimal location of the robot on the linear axis at which the seam is processed must be determined.
The objective of picking and sequencing the robot's trajectories is to find a time-optimal and collision-free production plan.
Such an optimal production plan may increase throughput, and automation of robot programming reduces the development time of new car bodies.

Quantum computers hold the promise to solve seemingly intractable problems across virtually all disciplines with chemistry and optimization being likely the first medium-term workloads. Specifically, the advent of quantum annealing devices such as the D-Wave Systems Inc.~quantum annealers \citep{johnson:11, bunyk:14, katzgraber:18a, hauke:20} has spawned an increased interest in the development of quantum-native, heuristic approaches to solve discrete optimization problems.
While impressive progress has been made over the last few years, the field is currently still in its infancy, but arguably at a transition point from mere academic research to industrialization. 
Currently, however, it is still unclear what type of quantum hardware and algorithms will deliver quantum advantage for a practical, real-world problem. Because of this, it is imperative to develop optimization methods that can bridge the gap until scalable quantum hardware is available, but also prepare customers to use specific optimization models that will eventually be able to run on quantum hardware.

The industry use case outlined above has been previously proposed as a potential industry reference benchmark problem for emerging quantum technologies~\citep{finzgar2022quark,luckow21, qutac}. 
To assess the capabilities of near-term quantum hardware and its potential impact for real-world industry use cases, here we follow a two-pronged approach.
On the one hand we provide and analyze small-scale numerical experiments on quantum annealing hardware (and hybrid extensions thereof), while on the other hand we design and implement a complementary nature-inspired solution strategy that can integrate quantum computing hardware into its larger framework and provide business value already today. 
Specifically, we put forward an end-to-end optimization pipeline that extends evolutionary meta-heuristics known as biased random-key genetic algorithms \citep{GonRes2011a} towards alternative algorithmic paradigms such as simulated annealing, in conjunction with model stacking and ensemble techniques for solution refinement. For automated hyperparameter tuning, we leverage Bayesian optimization techniques. 
By design, the resulting hybrid, quantum-ready solution is highly portable and should find applications across a myriad of industry-scale combinatorial optimization problems far beyond the use case studied in this work. 

This paper is structured as follows. 
In Sec.~\ref{preliminaries} we review the basic algorithmic concepts underlying our work, with details on biased random-key genetic algorithms, as well as quantum annealing. 
In Sec.~\ref{TheoreticalFramework} we then detail our theoretical framework, providing a comprehensive, quantum-ready optimization pipeline for solving robot path problems at industry scales. Section \ref{Numerics} describes systematic numerical benchmark experiments. 
Finally, in Sec.~\ref{Conclusion} we draw conclusions and give an outlook on future directions of research. \section{Preliminaries}
\label{preliminaries}

We start with a brief review of biased random-key genetic algorithms and dual annealing, to set notation and explain our terminology. Furthermore, we give a brief introduction to quantum annealing, as well as the quadratic unconstrained binary optimization (QUBO) formalism. 

\subsection{Biased random-key genetic algorithms}

Biased random-key genetic algorithms (BRKGA) \citep{GonRes2011a} represent a (nature-inspired, because genetic) heuristic framework for solving optimization problems.
It is a refinement of the random-key genetic algorithm of Bean \cite{Bea94a}.
While most of the work in the literature has focused on combinatorial optimization problems, BRKGA has also been applied to continuous optimization problems \cite{SilResPar2014a}.
The BRKGA formalism is based on the idea that a solution to an optimization problem can be encoded as a vector of random keys, i.e., a vector 
$\mathcal{X}$ in which each entry is a real number, generated at random in the interval $(0,1]$.
Such a vector $\mathcal{X}$ is mapped to a feasible solution of the optimization problem with the help of a \textit{decoder}, i.e.,~a deterministic algorithm that takes as input a vector of random keys and returns a feasible solution to the optimization problem, as well as the cost of the solution.

\subsubsection{BRKGA for traveling salesman and vehicle routing problems} 

The BRKGA framework is well suited for sequencing-type optimization problems, as relevant for our PVC use case. 
For example, consider the traveling salesman problem (TSP) where a salesman is required to visit $n$ given cities, each city only once, and do so taking a minimum-length tour.
A solution to the TSP is a permutation $\pi$ of the $n$ cities visited and its cost is 
\begin{displaymath}
c=\ell (\pi_1,\pi_2) + \ell (\pi_2,\pi_3) + \cdots + \ell (\pi_{n-1},\pi_n) + \ell (\pi_{n},\pi_1),
\end{displaymath}
where $\ell (i,j)$ is the distance between city $i$ and city $j$.
A possible decoder for the TSP takes the vector of random keys as input and sorts the vector in increasing order of its keys. The indices of the sorted vector make up $\pi$, the permutation of the visited cities.  As an example consider a TSP on five cities and let $\mathcal{X} = (0.45, 0.78, 0.15, 0.33, 0.95).$  The sorted vector is $s(\mathcal{X}) = (0.15, 0.33, 0.45, 0.78. 0.95)$ and its vector of indices is $\pi = (3, 4, 1, 2, 5)$ having cost
\begin{displaymath}
c=\ell (3,4) + \ell (4,1) + \ell (1,2) + \ell (2,5) + \ell (5,3).
\end{displaymath}
Consider now the vehicle routing problem (VRP) where we are given up to $p$ vehicles, a depot (node~0) and $n$ locations $\{1,2,\ldots,n\}$ that these vehicles must visit, starting and ending at the depot. Each location must be visited by exactly one vehicle and all locations must be visited. A solution to this problem is a set of $p$ permutations $\pi^1, \pi^2, \ldots, \pi^p$ such the $\pi^i \cap \pi^j = \varnothing$ (i.e., no two vehicles visit the same location),
for $i=1,\ldots,p-1$, $j = i +1, \ldots, p$ and $\cup_{i=1}^p \pi^i = \{1,2,\ldots,n\}$ (i.e., all locations are visited). In this solution $\pi^i$ indicates the sequence that vehicle $i$ will take.  
Suppose $\pi^1 = \{1, 3, 5\}$ and $\pi^2 = \{4, 2\}$, then vehicle~1 visits locations~1, 3, and~5, in this order, and vehicle~2 visits location~4 and then location~2.  
Both vehicles start and end their tours at node~0 (the depot).
The cost $C$ of this solution is the sum of the costs of the tours of each vehicle, i.e. $C=c^1+c^2$, where
\begin{displaymath}
c^1 = \ell (0,1) + \ell (1,3) + \ell (3,5) + \ell (5,0)
\end{displaymath}
and
\begin{displaymath}
c^2 = \ell (0,4) + \ell (4,2) + \ell (2,0).
\end{displaymath}
A possible decoder for the VRP takes as input a vector of $n+v$ random keys, sorts the keys in increasing order of their values, rotates the vector of sorted keys such that the largest of the $v$ keys is last in the array, and then uses the $v$ keys to indicate the division of locations traveled to by each vehicle.
For example, consider $n=5$ and $v=2$ and consider the following vector 
$\mathcal{X} = (0.45, 0.78, 0.15, 0.33, 0.95, \textbf{0.25}, \textbf{0.35})$
of random keys.
The first $n=5$ keys correspond to the locations to be visited by the $v=2$ vehicles.
The last two keys correspond to the two vehicles and are indicated in bold.
Sorting the keys in increasing order results in
$s(\mathcal{X}) = (0.15, \textbf{0.25}, 0.33, \textbf{0.35}, 0.45, 0.78,   0.95)$. The corresponding solution $(3, \mathbf{V_1}, 4, \mathbf{V_2}, 1, 2, 5)$ corresponds to the indices of the sorted
random-key vector.  For example, 3 is the index of the smallest key, 0.15, while 5 is the index of the largest key, 0.95.
$\mathbf{V_1}$ and $\mathbf{V_2}$ correspond, respectively, to the indices of
vehicle random keys in the sorted vector.
Rotating the elements of the solution vector circularly such that $V_2$ occupies the last position in the vector, we get $(1, 2, 5, 3, \mathbf{V_1}, 4, \mathbf{V_2})$ which translates into a solution where vehicle $V_1$ leaves the depot and visits locations 1, 2, 5, 3, and then returns to the depot and vehicle $V_2$ leaves the depot, visits location 4 and returns to the depot.  
The cost $C$ of this solution is the sum of the costs of the tours of each vehicle, i.\,e., $C=c^1+c^2$, where
\begin{displaymath}
c^1 = \ell (0,1) + \ell (1,2) + \ell (2,5) + \ell (5,3) + \ell (3,0)
\end{displaymath}
and
\begin{displaymath}
c^2 = \ell (0,4) + \ell (4,0).
\end{displaymath}

\begin{figure}
\includegraphics[width=1.0 \columnwidth]{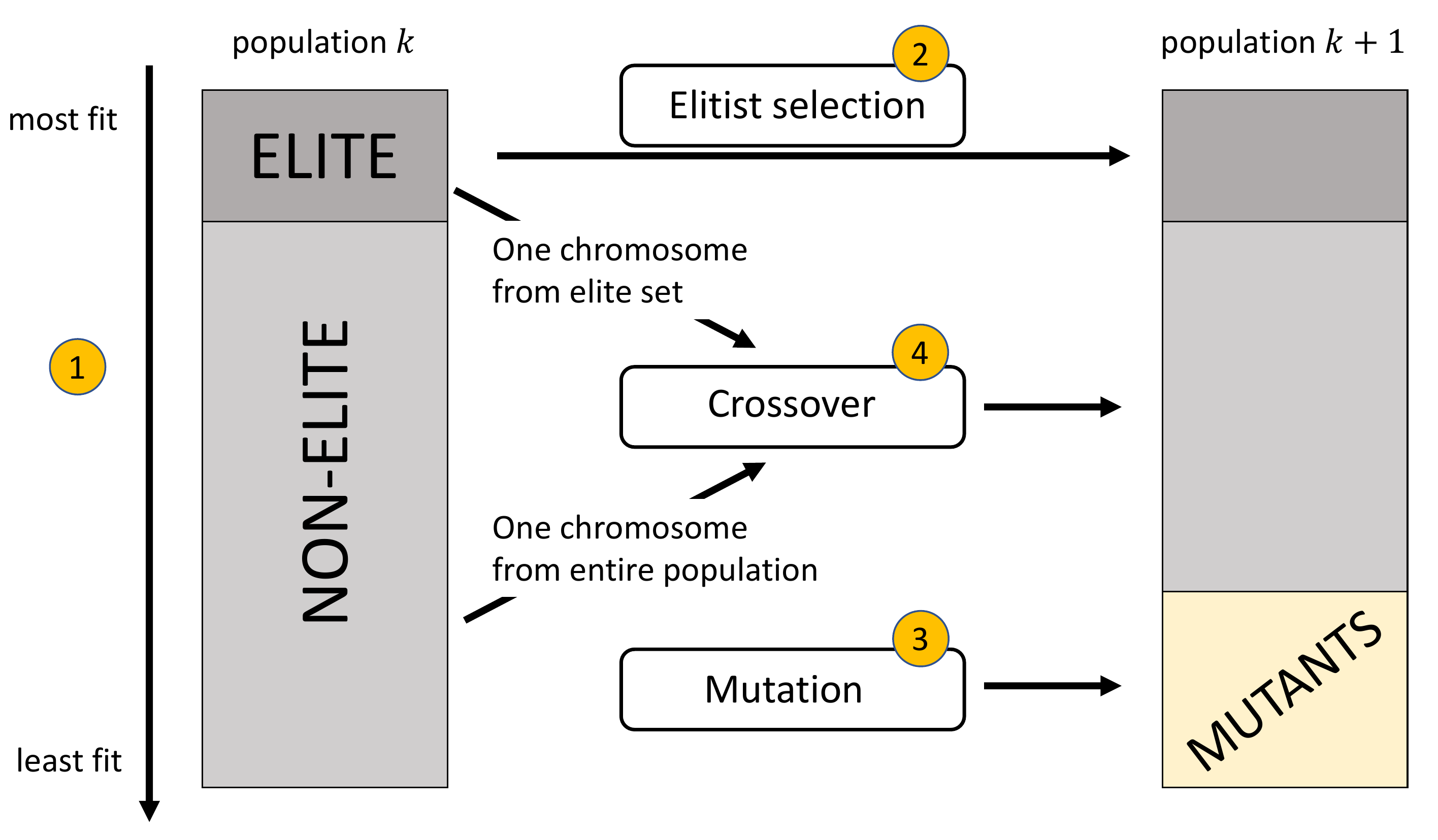}
\caption{(Color online) 
Schematic illustration of the evolutionary process underlying BRKGA. 
In step (1) the chromosomes within the current population $k$ are ranked according to their fitness values. 
In step (2) the elite individuals (those with the highest fitness scores) are copied over to population $k+1$. 
In step (3), for diversity and to combat local minima, new mutant individuals are randomly generated and added to the population $k+1$. 
In step (4) the remaining portion of the population $k+1$ is topped up with offsprings generated by a biased crossover that mates elite with nonelite parents.
\label{fig:brkga}
}
\end{figure}

\subsubsection{Anatomy of BRKGA} 

BRKGA starts with an initial population $\mathcal{P}_0$ of $p$ random-key vectors, each of length $N$.
A decoder is applied to each vector to produce a solution to the problem being solved.
Over a number of generations, BRKGA evolves this population until some stopping criterion is satisfied. Populations $\mathcal{P}_1, \mathcal{P}_2, \ldots, \mathcal{P}_K$ are generated over $K$ generations. The best solution in the final population is output as the solution of the BRKGA.
BRKGA is an elitist algorithm in the sense that it maintains an elite set $\mathcal{E}$
with the best solutions found during the search. The dynamics of the evolutionary process is simple. The population $\mathcal{P}_k$ of each generation is made up of two sets of random vectors, the elite set and the remaining solutions, the nonelite set. 
To generate population $\mathcal{P}_{k+1}$ from $\mathcal{P}_{k}$ the elite set of $\mathcal{P}_{k}$ is copied, without modification to $\mathcal{P}_{k+1}$.
This accounts for $p_e = |\mathcal{E}|$ elements.
Next, a set $\mathcal{M}$ of mutant solutions (randomly generated random-key vectors) is generated and added to $\mathcal{P}_{k+1}$.
This accounts for an additional $p_m = |\mathcal{M}|$ elements.
The remaining $p - p_e - p_m$ elements of $\mathcal{P}_{k+1}$ are generated through parameterized uniform crossover \cite{SpeDej91}.
Two parents are selected at random, with replacement, one from the elite set of $\mathcal{P}_{k}$, the other from the nonelite set.
Denote these parents, respectively, as the elite parent $\mathcal{X}_a$ and the nonelite
parent $\mathcal{X}_b$.
The offspring $\mathcal{X}_c$ is generated as follows.  For $i=1,\ldots,N$, let $\mathcal{X}_c[i] \gets \mathcal{X}_a[i]$ with probability $\Pi > \frac{1}{2}$.  Else, with probability $1 - \Pi$, $\mathcal{X}_c[i] \gets \mathcal{X}_b[i]$.
Offspring $\mathcal{X}_c$ is added to $\mathcal{P}_{k+1}$.
This process is repeated until all $p - p_e - p_m$ offsprings are added to $\mathcal{P}_{k+1}$, completing a population of $p$ elements.
The $p_e$ random-key vectors with the overall best solutions of $\mathcal{P}_{k+1}$ are placed in the population's elite set, while the remaining vectors are nonelite solutions.  
A new iteration starts by setting $k \gets k+1$. 
For illustration, the evolutionary process underlying BRKGA is illustrated schematically in Fig.~\ref{fig:brkga}.

\subsubsection{BRKGA in action} 

BRKGA is a general-purpose optimizer where only the decoder needs to be tailored towards a particular problem.  
In addition, several hyperparameters need to be specified.
These are limited to the length $N$ of the vector of random-keys, the size $p$ of the population, the size of the elite set $p_e < p/2$, the size of the set of mutants $p_m \leq p - p_e$, and the probability $\Pi > 1/2$ that the offspring inherits the keys of the elite parent.
In addition, a stopping criterion needs to be given.  
That can be, for example, a maximum number of generations, a maximum number of generations without improvement, a maximum running time, or some other criterion.
Several application programming interfaces (API) have been proposed for BRKGA (\cite{TosRes2015a},\cite{AndTosGonRes21a}), including some based on C++, Python, Julia, and Java.  
With these APIs, the user needs only to define a decoder and specify the hyperparameters of the algorithm.

\subsubsection{Extensions for BRKGA} 

It should be noted that several extensions have been proposed for BRKGA.  Decoding can be done in parallel \cite{ResTosGonSil2012a}.
Instead of evolving a single population, several populations can be evolved in an island model \cite{GonRes2012a}.
Restarts are known to improve the performance of stochastic local search optimization algorithms \cite{LubSinZuc1993a}. The number of generations without improvement can be used to trigger a restart in a BRKGA where the current population is replaced by a population of $p$ vectors of random keys.  In BRKGA with restart \cite{ResMorGonTos13a} a maximum number of restarts can be used as a stopping criterion. Instead of mating two parents, mating can be done with multiple parents \cite{LucAndResMiy2014a}.
Finally, path-relinking strategies can be applied in the space of random keys as a problem-independent intensification operator \cite{AndTosGonRes21a}.

\subsection{Dual Annealing}

Dual Annealing (DA) is a stochastic, global (nature-inspired) optimization algorithm. 
Here, we provide a brief overview of the DA algorithm as used in our extension of the random key optimizer (RKO), described in more depth in Section \ref{TheoreticalFramework}. 
We use the DA implementation provided in the SciPy library \cite{scipy:2020}. 
This implementation is based on Generalized Simulated Annealing (GSA), which generalizes classical simulated annealing (CSA) and the extended fast simulated annealing (FSA) into one unified algorithm \cite{tsallis:96, xiang:97}, coupled with a strategy for applying a local search on accepted locations for further solution refinement. 
GSA uses a modified Cauchy-Lorentz visiting distribution, whose shape is controlled by the visiting parameter $q_{v}$
\begin{equation}
g_{q_{v}}(\Delta \mathcal{X}(t)) \propto 
\frac{[T_{q_{v}}(t)]^{\frac{-D}{3-q_{v}}}} 
{[1+(q_{v}-1) \frac{(\Delta \mathcal{X}(t))^{2}}{[T_{q_{v}}(t)]^{\frac{2}{3-q_{v}}}}]^{\frac{1}{q_{v}-1} + \frac{D-1}{2}}}, \label{eq:DA-visiting}
\end{equation}
where $t$ is the artificial time (algorithm iteration). 
This distribution is used to generate a candidate jump distance $\Delta \mathcal{X}(t)$ under temperature $T_{q_{v}}$, which is the step from variable $\mathcal{X}(t)$ the algorithm proposes to take. 
If this proposed step yields an improved cost, it is accepted. 
If the step does not improve the cost, it may be accepted with acceptance probability $p_{q_{a}}$
\begin{equation}
p_{q_{a}} = \mathrm{min}\{1, \mathrm{max}\{0, [1 - (1 - q_{a}) \beta \Delta E]^{\frac{1}{1-q_{a}}} \} \}, \label{eq:DA-acceptance}
\end{equation}
where $\Delta E$ is the change in energy (cost) of the system, $q_{a}$ is an algorithm hyperparameter, and $\beta \equiv 1/(\kappa T_{q_{v}}(t))$ refers to the inverse temperature, with Boltzmann constant $\kappa$.
If the proposed step is accepted, this yields an update step of
\begin{equation}
\mathcal{X}(t) = \mathcal{X}(t-1) + \Delta \mathcal{X}(t), \label{eq:DA-update}
\end{equation}
otherwise, $\mathcal{X}(t)$ remains unchanged. 
The artificial temperature $T_{q_{v}}(t)$ is decreased according to the annealing schedule  
\begin{equation}
T_{q_{v}}(t) = T_{q_{v}}(1) \frac{2^{q_{v}-1} -1}{(1+t)^{q_{v}-1}-1}, \label{eq:DA-temp}
\end{equation}
where $T_{q_{v}}(1)$ is the starting temperature, with default $T_{q_{v}}(1)=5230$. 
As the algorithm runs through this parameterized annealing schedule, both acceptance probabilities $p_{q_{a}}$ as well as jump distances $\Delta \mathcal{X}(t)$ decrease over time; this has been shown to yield improved global convergence rates over FSA and CSA \cite{xiang:13}.

After each GSA temperature step, a local search function is invoked, which in the bounded variable case (as ours is here, i.e. $\mathcal{X}(t) \in [0, 1]^{n}$) defaults to the L-BFGS-B algorithm. 
At each iteration, the L-BGFS-B algorithm runs a line search along the direction of steepest gradient descent around $\mathcal{X}(t)$ while conforming to provided bounds.
For more details, see Ref.~\cite{byrd:95}.
Once the local search converges (or exits, i.e., by reaching invocation limits), the found solution $\mathcal{X}(t)$ is used as the starting point for the next step in the GSA algorithm.

This dual annealing process of GSA followed by L-BGFS-B runs until convergence, or until the algorithm exits due to maximum iterations, as set by the algorithm's hyperparameter \texttt{maxiter}.
If the artificial temperature $T_{q_{v}}(t)$ shrinks to a value smaller than $\mathcal{R}*T_{q_{v}}(1)$ (with corresponding hyperparameter \texttt{restart\_temp\_ratio}), then the dual annealing process is restarted, with temperature reset to $T_{q_{v}}(1)$ and a random (bounded) position is provided for $\mathcal{X}_{0}$. 
Note that the algorithm iteration counts are not reset in this case, so overall algorithm runtime remains tractable.

\subsection{Quantum Annealing and the QUBO formalism}

\begin{figure}
\includegraphics[width=1.0 \columnwidth]{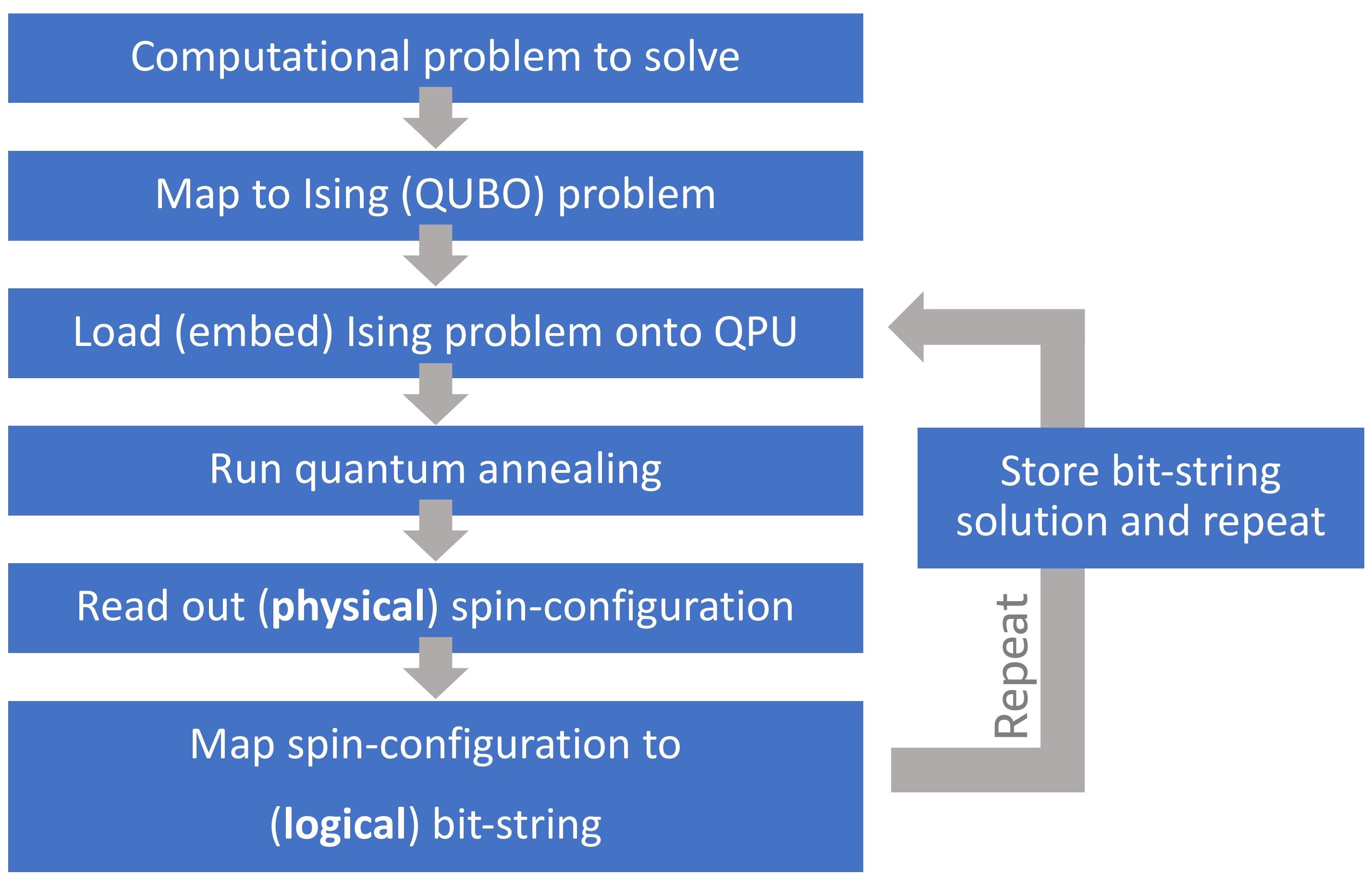}
\caption{(Color online) 
Flow chart illustrating the end-to-end workflow for solving a combinatorial optimization problem on a quantum annealer.
First, the problem has to be cast as a QUBO (or equivalently Ising) Hamiltonian.
This abstract QUBO problem is then mapped onto the physical QPU, typically at the expense of an enlarged number of variables (given the sparse connectivity of the underlying quantum chip). 
Next, quantum annealing is used to find a high-quality variable configuration. 
This solution is mapped back to a bit-string (of logical variables) corresponding to a solution of the original optimization problem. 
Given the probabilistic nature of this process, it is typically repeated multiple times, followed by a statistical analysis.
The goal is to find a configuration of variables that (approximately) minimizes the objective function. 
Further details are provided in the main text. 
\label{fig:qa}
}
\end{figure}

Quantum computers are devices that harness quantum phenomena not available to conventional (classical) computers. 
Today, the two most prominent paradigms for quantum computing involve (universal) circuit-based quantum computers, and (special-purpose) quantum annealers \citep{hauke:20}.  
While the former hold the promise of exponential speed-ups for certain problems, in practice circuit-based devices are extremely challenging to scale up, with current quantum processing units (QPUs) providing about one hundred (\textit{physical}) qubits \citep{hauke:20}. 
Moreover, to fully unlock any exponential speed-up, perfect (\textit{logical}) qubits have to be realized, as can be done using quantum error correction, albeit with a large overhead, when encoding one logical (noise-free) qubit in many physical (noisy) qubits.  
Conversely, quantum annealers are \textit{special-purpose} machines designed to solve certain combinatorial optimization problems belonging to the class of Quadratic Unconstrained Optimization (QUBO) problems. 
Since quantum annealers do not have to meet the strict engineering requirements that universal gate-based machines have to meet, already today this technology features $\sim 5000$ (physical) analog superconducting qubits.

\subsubsection{QUBO formalism} 

Recently, the QUBO framework has emerged as a powerful approach that provides a common modelling framework for a rich variety of NP-hard combinatorial optimization problems \citep{lucas:14, kochenberger:14, glover:18, anthony:17}, albeit with the potential for a large variable overhead for some use cases.
Prominent examples include the maximum cut problem, the maximum independent set problem, the minimum vertex cover problem, and the traveling salesman problem, among others \citep{lucas:14, kochenberger:14, glover:18}. The cost function for a QUBO problem can be expressed in compact form with the Hamiltonian

\begin{equation}
H_\mathrm{QUBO} = \mathbf{x}^{\intercal}Q\mathbf{x} = \sum_{i,j} x_{i}Q_{ij}x_{j}, \label{eq:H-QUBO}
\end{equation}
where $\mathbf{x}=(x_{1}, x_{2}, \dots)$ is a vector of binary decision variables and the QUBO matrix $Q$ is a square matrix 
that encodes the actual problem to solve. 
Without loss of generality, the $Q$-matrix can be assumed to be symmetric or in upper triangular form \citep{glover:18}. 
We have omitted any irrelevant constant terms, as well as any linear terms, as these can always be absorbed into the $Q$-matrix because $x_{i}^2=x_{i}$ for binary variables $x_{i} \in \{0,1\}$.  
Problem constraints---which relevant for many real-world optimization problems---can be accounted for with the help of penalty terms entering the objective function, as detailed in Ref.~\citep{glover:18}. 
Explicit examples will be provided below. 

The significance of QUBO formalism is further illustrated by the close relation to the famous \textit{Ising} model \cite{ising:25}, which is known to provide mathematical formulations for many NP-complete and NP-hard problems, including all of Karp's 21 NP-complete problems \citep{lucas:14}. 
As opposed to QUBO problems, Ising problems are described in terms of binary (classical) spin variables $z_{i}\in\{-1,1\}$, that can be mapped straightforwardly to their equivalent QUBO form, and vice versa, using $z_{i}=2x_{i}-1$. 
The corresponding classical Ising Hamiltonian reads 
\begin{equation}
H_\mathrm{Ising} = -\sum_{i,j} J_{ij}z_{i}z_{j} - \sum_{i}h_{i}z_{i}, \label{eq:H-Ising}
\end{equation}
with two-body spin-spin interactions $J_{ij}=-Q_{ij}/4$, and local fields $h_{i}$ (note that a trivial constant has been omitted).
If the couplers $J_{ij}$ are chosen from a random distribution, the Ising model given above is also known as a \textit{spin glass}. 
By definition, both the QUBO and the Ising models are quadratic in the corresponding decision variables. 
If the original optimization problem involves $k$-local interactions with $k>2$, degree reduction schemes have to be involved, at the expense of the aforementioned overhead in terms of the number of variables \citep{hauke:20}. In general, one disadvantage of solving problems in a QUBO formalism on quantum annealing hardware lies in the fact that the problem has to be first mapped to a binary representation, then locality has to be reduced to $k \le 2$ (see below for details). 

\subsubsection{Quantum annealing} 

Quantum annealing (QA) \cite{kadowaki:98,farhi:01} is a metaheuristic for solving combinatorial optimization problems on special-purpose quantum hardware, as well as via software implementations on classical hardware using quantum Monte Carlo \cite{isakov:16}.
In this approach the solution to the original optimization problem is encoded in the ground state of the so-called problem Hamiltonian $\hat{H}_{\mathrm{problem}}$.
Finding the optimal assignment $\mathbf{z}^{*}$ for the classical Ising model \eqref{eq:H-Ising} is equivalent to finding the ground state of the corresponding problem Hamiltonian, where we have promoted the classical spins $\{z_{i}\}$ to quantum spin operators $\{\hat{\sigma}_{i}^{z}\}$, also known as Pauli matrices, thus describing a collection of interacting qubits.  
To (approximately) find the \textit{classical} solution $\{z_{i}^{*}\}$, quantum annealing devices then undergo the following protocol: Start with the algorithm by initializing the system in some easy-to-prepare ground state of an initial Hamiltonian $\hat{H}_{\mathrm{easy}}$, which is chosen to not commute with $\hat{H}_{\mathrm{problem}}$. 
Following the adiabatic approximation \citep{hauke:20}, the system is then slowly annealed towards the so-called problem Hamiltonian $\hat{H}_{\mathrm{problem}}$,
whose ground state encodes the (hard-to-prepare) solution of the original optimization problem. 
This is commonly done in terms of the annealing parameter $\tau$, defined as $\tau = t/T_{A} \in [0,1]$, where $t$ is the physical wall-clock time, and $T_{A}$ is the annealing time. 
In the course of this anneal, ideally, the probability to find a given classical configuration converges to a distribution that is strongly peaked around the ground state of $H_{\mathrm{Ising}}$. 
Overall, the protocol is captured by the time-dependent Hamiltonian 

\begin{equation}
\hat{H}(\tau) = A(\tau) \hat{H}_{\mathrm{easy}}+ B(\tau) \hat{H}_{\mathrm{problem}}, \label{eq:H-annealing}
\end{equation}
where the functions $A(\tau), B(\tau)$ describe the annealing schedule, with $A(0)/B(0) \gg 1$, and $A(1)/B(1) \ll 1$.
For example, a simple, linear annealing schedule is given by $A(\tau)=1-\tau$, and $B(\tau) = \tau$.
Because of manufacturing constraints, current experimental devices can only account for $2$-local interactions, with a cost function described by 

\begin{equation}
\hat{H}_\mathrm{problem} = -\sum_{i,j} J_{ij}\hat{\sigma}_{i}^{z}\hat{\sigma}_{j}^{z} - \sum_{i}h_{i}\hat{\sigma}_{i}^{z}, \label{eq:H-problem}
\end{equation}
while the initial Hamiltonian is typically chosen as 

\begin{equation}
\hat{H}_\mathrm{easy} = -\sum_{i} \hat{\sigma}_{i}^{x}, \label{eq:H-easy}
\end{equation}
that is a transverse-field driving Hamiltonian responsible for quantum tunneling between the classical states making up the computational basis states. 
Because the final Hamiltonian \eqref{eq:H-problem} only involves commuting operators $\{\hat{\sigma}_{i}^{z}\}$, the final solution $\{z_{i}^{*}\}$ can be read out as the state of the individual qubits via a measurement in the computational basis.

\subsubsection{Embedding} 

Solving an optimization problem of QUBO form on QA hardware, however, frequently involves one more step, typically referred to as \textit{embedding} \citep{choi:08}. 
Because of manufacturing constraints, today's quantum annealers based on superconducting technology only come with limited connectivity, i.e., not every qubit is physically connected to every other qubit. 
In fact, typically the on-chip matrix $J_{ij}$ is sparse.  
If, however, the problem's required logical connectivity does not match that of the underlying hardware, one can effectively replicate the former using an embedding strategy by which several \textit{physical} qubits are combined into one \textit{logical} qubit. 
The standard approach to do so is called \textit{minor embedding} which provides a mapping from a (logical) graph to a sub-graph of another (hardware) graph.  
One can then solve high-connectivity problems directly on the sparsely connected chip, by sacrificing physical qubits accordingly to the connectivity of problem, typically introducing a considerable overhead with multiple physical qubits making up one logical variable. This limitation makes the problems subsequently harder to solve than in their native formulation \cite{zaribafiyan:17}.  
Specifically, in the extreme case of a fully-connected graph (as relevant for traveling salesman problem) only approximately $\sim 64$ \textit{logical} spin variables can be embedded within the D-Wave 2000Q quantum annealer that nominally features $\sim 2000$ \textit{physical} qubits. Another example, circuit fault diagnosis, is analyzed in detail in Ref.~\citep{perdomo:17y} and further highlights some of these limitations. Finally, when including constrained problems, the coupler distributions tend to broaden, which, in turn, results in an additional disadvantage due to the limited precision of the analog device.
 
\subsubsection{QA workflow} 

Today, quantum annealing devices as provided by D-Wave Systems Inc.~can be conveniently accessed through the cloud. 
In Fig.~\ref{fig:qa} we detail the end-to-end workflow for solving some combinatorial optimization problem on such a quantum annealer from a practitioner's point of view. 
Below, we will put this workflow into practice when solving small instances of our use case on D-Wave, as available on Amazon Braket, and hybrid extensions thereof. \section{Theoretical Framework}
\label{TheoreticalFramework}

\begin{figure*} [h]
\includegraphics[width=2.0 \columnwidth]{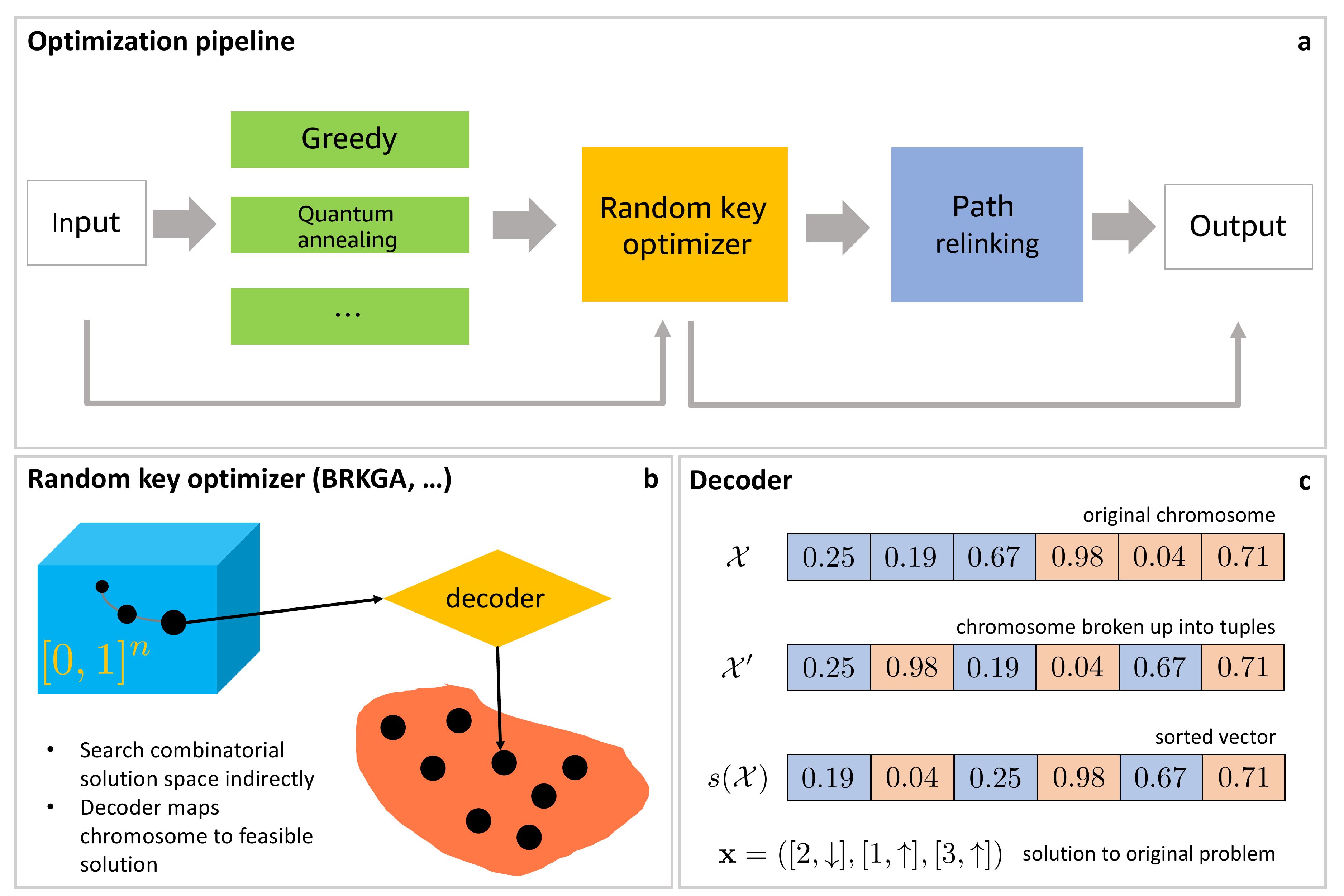}
\caption{(Color online) 
Schematic illustration of our approach. 
\textbf{(a)} Flow chart of our end-to-end optimization pipeline. 
The core routine takes the problem input, here specified in terms of cost values associated with pairs of nodes, and feeds this input into the random key optimizer (RKO).
The latter heuristically searches for an optimized tour of composite nodes, which represents the pipeline's output. 
This core routine can be extended with additional upstream and downstream modules for further solution refinement. Upstream solutions provided by alternative algorithms (such as greedy algorithms, quantum annealing, etc.) can be used to warm-start the RKO. 
Akin to ensemble techniques routinely used in machine learning, a pool of different solutions may help identify high-quality solutions, adding high-quality diversity into the initial population of the RKO module. 
Downstream, further solution refinement can be achieved through path-relinking techniques, by forming superpositions of high-quality solution candidates. 
\textbf{(b)} Schematic illustration of the random key optimizer (RKO). 
The key characteristic of the RKO is a clear separation between problem-independent modules (as illustrated by the blue hypercube that hosts the chromosome $\mathcal{X}$) and the problem-dependent, deterministic decoder that maps $\mathcal{X}$ to a solution of the original problem with associated cost (or fitness) value. 
By design our decoder ensures that problem constraints (e.g., every node has to be visited exactly once) are satisfied. 
In BRKGA the trajectory of the chromosome $\mathcal{X}$ is set by evolutionary principles, but generalizations to alternative algorithmic paradigms such as simulated annealing are straight-forward. \textbf{(c)} Example illustration of the decoding of a chromosome $\mathcal{X}$, made of random keys in $(0,1]$, into a solution to the original combinatorial optimization problem. 
We consider a sequencing problem paired with a binary decision variable (such as the binary direction variable $d=\uparrow, \downarrow$) for $n=3$ composite nodes.
The first block of the chromosome (highlighted in blue) encodes the solution to the sequencing problem, while the second block (highlighted in red) encodes the additional binary variable, thus representing a minimal example for the concept of a \textit{composite} node. 
The chromosome can be broken up into $n$ tuples, one encoding a single node each. 
The decoder then performs simple sorting according to the first entry in the tuple, yielding the sorted vector $s(\mathcal{X})$. 
Finally, the solution to the original problem $\mathbf{x}$ is given by the indices of this sorting routine paired with a simple threshold routine applied to the tuples' second entry. Here, the threshold is set at $0.5$. 
Further details are provided in the main text. 
\label{fig:framework}
}
\end{figure*}

In this section we discuss in detail the theoretical framework underlying our work, as outlined in Fig.~\ref{fig:framework}.
We first define notation, introducing the concept of an abstract, composite \textit{node} that encapsulates information about the seam number together with other relevant degrees of freedom (such as tool configuration or position of the robot). 
Next we detail our classical as well as quantum-native solution strategies to the PVC use case. 
Specifically, we show how BRKGA can be applied to the PVC use case, by proposing an efficient decoder design that natively accounts for the problem's constraints.
Finally, we detail how this use case can be described within the quantum-ready QUBO framework. 

A composite \textit{node} can be viewed as a generalization of a \textit{city} in the canonical TSP.
In analogy of the TSP, the goal is to identify an optimal sequence of nodes, with a node encoding not only spatial information, but also other categorical features relevant to the use case at hand. Specifically, in our setup we define a node as a quintuple of the form
\begin{equation}
\mathrm{node} = [s, d, t, c, p]. \label{eq:node}
\end{equation}
Generalizations to other problems are straightforward. 
Here, a node encapsulates information about the seam index $s=1, \dots, n_{\mathrm{seams}}$, the direction $d=0,1$ by which a given seam is sealed, the tool $t=1, \dots, n_{\mathrm{tools}}$ and tool configuration $c=1, \dots, n_{\mathrm{config}}$ used, as well as the (discretized) linear-axis position $p=1, \dots, n_{\mathrm{position}}$. 
This definition of a generalized, composite node accounts for the fact that (i) any seam can be sealed in one of two directions, (ii) a robot can seal a given seam using one of several tools, (iii) which (at the same time) can be employed in different configurations, and (iv) a robot can take one of several positions along a fixed rail.
All coordinates can be described by integer values. 
The problem is then specified in terms of cost values $w_{\mathrm{node}_{i}}^{\mathrm{node}_{j}}$ (in seconds) for the robot to move from one of the endpoints of $\mathrm{node}_{i}$ to one of the endpoints of $\mathrm{node}_{j}$, including both the cost associated with applying PVC to $\mathrm{node}_{i}$ as well as proceeding in idle mode to $\mathrm{node}_{j}$. 
As illustrated in Fig.~\ref{fig:scheme}, every seam has two endpoints, i.e., a degree of freedom captured by the binary direction variable $d=0,1$.
For illustration, a sample data set is shown in Tab.~\ref{tab:sample-data}. 
For industry-relevant problem instances such a data set has roughly one million rows, only providing preselected, \textit{feasible} connections, as (in practice) many node pairs represent infeasible robot routes because of obstacles. 
Finally, we note that each robot has a home (or base) position from which it starts its operation, and where its tour comes to an end; this home position is associated with the node $[0,0,0,0,0]$. 
More details on the generation of real-world data-sets will be given in Sec.~\ref{Numerics}. 

\begin{table}
\caption{
Example data set (cost matrix) for illustration. 
The problem is specified in terms of cost values (in seconds) for pairs of nodes, with every node described by a tuple $[s, d, t, c, p]$ that captures the seam index $s$, direction $d$, tool $t$, tool configuration $c$, and robot position $p$. 
Further details are provided in the main text. 
\label{tab:sample-data}
}
\begin{tabular*}{\columnwidth}{@{\extracolsep{\fill}} c c c c c | c c c c c | c }
\hline
\hline
 \multicolumn{5}{c}{node (from)} & \multicolumn{5}{c}{node (to)} & cost \\
 \hline
 $s$ & $d$ & $t$ & $c$ & $p$ & $s$ & $d$ & $t$ & $c$ & $p$ & $w$ [s] \\ [0.5ex] 
 \hline\hline
 0 & 0 & 0 & 0 & 0 & 18 & 1 & 1 & 1 & 1 & 0.877 \\ 
 11 & 2 & 1 & 0 & 1 & 12 & 1 & 2 & 0 & 1 & 0.473\\  
 11 & 2 & 1 & 0 & 1 & 12 & 0 & 3 & 0 & 0 & 0.541\\  
 32 & 2 & 1 & 2 & 1 & 25 & 2 & 1 & 2 & 1 & 0.558\\
 \vdots &  &  &  &  & \vdots &  &  &  &  & \vdots\\[1ex] 
\hline
\hline
\end{tabular*}
\end{table}

\subsection{Robot motion planning with nature-inspired algorithms}

We now discuss our end-to-end optimization pipeline for robot motion planning, as schematically depicted in Fig.~\ref{fig:framework}. 
We first detail the core optimization routine, dubbed random-key optimizer (RKO), with a focus on the use-case specific decoder design, before discussing potential upstream and downstream extensions for further solution refinement.
Specifically, with the RKO concept, we introduce a generalization of the BRKGA formalism and its distinct separation of problem-independent and problem-dependent modules towards alternative optimization paradigms such as simulated annealing.
More details are provided below. 


\subsubsection{Core pipeline: Decoder design} 

We now detail an example decoder design, as used in our numerical experiments. 
The problem input is given in terms of a generalized cost matrix as displayed in Tab.~\ref{tab:sample-data}. 
Similar to the TSP, our goal is to identify a minimum-cost tour (as a sequence of \textit{nodes}) that visits $n$ given \textit{seams}, each seam only once, while specifying the additional degrees of freedom making up a composite \textit{node}.  
As illustrated in Fig.~\ref{fig:framework}(b) the \textit{decoder} plays an integral part in our random-key approach. 
While the problem \textit{encoding} is specified by the evolutionary part underlying BRKGA, \textit{decoding} is controlled through the design of the decoder.
Here, the decoder is designed as follows (see Appendix \ref{decoder} for implementation details in python code). 
The decoder takes a vector $\mathcal{X}$ of $N=\mathcal{D} \cdot n_{\mathrm{seams}}$ random keys as input, sorts the keys associated with the seam numbers $s$ in increasing order of their values, and applies simple thresholding logic to the remainder of keys; 
see Fig.~\ref{fig:framework}(c) for an illustration. 
In our case the number of features $\mathcal{D}$ is $\mathcal{D}=5$. 
Similar to the TSP example outlined above, the indices of the sorted vector component make up $\pi$, the permutation of the visited seams; see the blue block in Fig.~\ref{fig:framework}(c). 
As opposed to the TSP, however, we have to assign discrete values for the remaining node degrees of freedom as well, as shown in the red block in Fig.~\ref{fig:framework}(c). 
For example, if the corresponding original variable is binary, as is the case in Fig.~\ref{fig:framework}(c), the thresholding logic reduces to $\mathrm{int}(\mathcal{X}_{i}) \in \{0,1\}$ but generalizations to variables with larger cardinality are straightforward. 
For example, if a larger cardinality is assumed for the original variable $\mathcal{Y}_i$, say $\mathcal{Y}_i \in \{ 1, 2, \ldots, C\}$, then $\mathcal{Y}_i = k$ if $\mathcal{X}_i \in ( \frac{k-1}{C},\frac{k}{C}]$, for $k=1,2,\ldots,C$.
Note that our mathematical representation by design generates feasible routes only where every seam is visited exactly once, while only scaling linearly with the number of seams $\sim n_{\mathrm{seams}}$, and with a prefactor set by the number of degrees of freedom. 
While the original cost matrix as displayed in Tab.~\ref{tab:sample-data} features feasible connections only, the decoder may suggest infeasible moves that have been preselected from the original data set. 
By padding the cost matrix with prohibitively large cost values for these types of moves, over the course of the evolution the algorithm will learn to steer away from these bad-fitness solutions.  As detailed in Sec.~\ref{Numerics} numerically, we find that our solution always arrives at feasible, low-cost tours that include feasible moves only. 

\subsubsection{Algorithmic generalizations} 

In the common BRKGA framework the trajectory of every chromosome $\mathcal{X}$ in the abstract half-open hypercube of dimension $(0,1]^{n}$ is governed by evolutionary principles, as detailed in Sec.~\ref{preliminaries}. 
However, alternative algorithmic paradigms such as simulated annealing (SA) \cite{kirkpatrick:83} and related methods can be readily used as well, all within our random-key formalism. 
That is because for any chromosome $\mathcal{X}$ the decoder does not only provide the decoded solution $s(\mathcal{X})$ but also the fitness (or cost) value $f(\mathcal{X})$, in our case defined as the total cost of the tour.
Black-box query access to $f(\mathcal{X})$, however, is sufficient for optimization routines such as SA to perform an update on the solution candidate $\mathcal{X}$ and continue with training till some (algorithm-specific) stopping criterion is fulfilled. 
To illustrate this point, we have performed numerical experiments based on the dual annealing algorithm \citep{xiang:97}. 
As detailed in Sec. \ref{preliminaries}, this stochastic approach combines classical SA with local search strategies for further solution refinement. 
We will refer to this annealing-based extension of the random-key formalism as RKO-DA. 
Numerical results and more details are presented in Sec.~\ref{Numerics}.

\subsubsection{Warmstarting} 

The core optimization routine outlined above can be extended with additional upstream logic. 
Specifically, in analogy to model-stacking techniques commonly used in machine-learning pipelines, solutions provided by alternative algorithms (such as linear programming, greedy algorithms, quantum annealing, etc.) can be used to warm-start the RKO, as opposed to cold-starts with a random initial population $\mathcal{P}_{0}$.
Similar to standard ensemble techniques, the output of several optimizers may be used to seed the input for RKO, thereby leveraging information learned by these while injecting diverse quality into the initial solution pool $\mathcal{P}_{0}$. 
By design this strategy can only improve upon the solutions already found, as elite solutions are not dismissed and just propagate from one population to the next in the course of the evolutionary process. 
The technical challenge is to invert the decoder with its inherent many-to-one mapping. 
To this end we propose the following randomized heuristic.
Consider a given permutation $\pi$ such as $\pi=(4,2,3,1)$, with $n=4$. 
Our goal is then to design an algorithm which produces a random key $\mathcal{X}$ that (when passed to the decoder) is decoded to the permutation $\pi$. 
To this end we chop up the interval $(0,1]$ into evenly spaced chunks of size $\Delta=1/n=0.25$, with centers $\bar{\mathcal{X}}_{i}$ at $0.125$, $0.375$, $0.625$, and $0.875$.
We then loop through the input sequence $\pi$ and assign these center values to appropriate positions in $\mathcal{X}$, as $\mathcal{X}=(\cdot,\cdot,\cdot,\cdot) \rightarrow (\cdot,\cdot,\cdot,0.125) \rightarrow (\cdot,0.375,\cdot,0.125) \rightarrow \dots \rightarrow (0.875,0.375,0.625,0.125)$.
When sorting this key we obtain the desired sequence of indices given by $\pi=(4,2,3,1)$. 
Finally, we randomize this deterministic protocol by adding uniform noise $\delta_{i} \in (\bar{\mathcal{X}}_{i}-\Delta/2, \bar{\mathcal{X}}_{i}+\Delta/2)$ to each element in $\mathcal{X}$, thus providing a randomized chromosome such as $\mathcal{X}=(0.93, 0.31, 0.67, 0.08)$.
By repeating the last step $m$ times, one can generate a pool of $m$ warm chromosomes. 
For the remaining categorical features $d, t, c, p$ it is straightforward to design a similar randomized protocol. 
For example, for the binary feature $d$ we generate a random number in $(0,0.5]$ if $d=0$, and a random number in $(0.5,1]$ if $d=1$. 

\subsubsection{Path relinking} 

Our core pipeline can be further refined with path-relinking (PR) strategies for potential solution refinement. 
Several PR strategies are known, some of them operating in the space of random keys (also known as implicit PR \cite{AndTosGonRes21a}) for problem-independent intensification, and some of them operating in the decoded solution space. 
The common theme underlying all PR approaches is to search for high-quality solutions in a space spanned by a given pool of elite solutions, by exploring
trajectories connecting elite solutions \cite{glover:97, resende:10}; one or more paths in the search space graph connecting these solutions are explored in the search for better solutions.
In addition to these existing approaches, here we propose a simple physics-inspired PR strategy that can be applied post-training. 
Consider two high-quality chromosomes labelled as $\mathcal{X}_{1}$ and $\mathcal{X}_{2}$, respectively. 
We can then heuristically search for better solutions with the hybrid (superposition) Ansatz 
\begin{equation}
\mathcal{X}(\alpha) = (1-\alpha) \mathcal{X}_{1}+ \alpha \mathcal{X}_{2}, \label{eq:PI-relinking}
\end{equation}
with $\alpha \in [0,1]$. 
We then scan the parameter $\alpha$ and query the corresponding fitness by invoking the decoder (without having to run the RKO routine again).
For $\alpha=0$ and $\alpha=1$ we recover the existing chromosomes $\mathcal{X}_{1}$ and $\mathcal{X}_{2}$, respectively, but better solutions may be found along the trajectory sampled with the hybridization parameter $\alpha$.   

\subsubsection{Restarts} 

Finally, restart strategies as described in Sec.~\ref{preliminaries} can be readily integrated into our larger optimization pipeline. Numerical experiments including restarts are detailed in Sec.~\ref{Numerics}. 

\subsection{Robot motion planning with quantum-native algorithms}

We now detail a quantum-native QUBO formulation for our industry use case, followed by resource estimates for the number of logical qubits required to solve this use case at industry-relevant scales.

\subsubsection{QUBO representation} 

We introduce binary (one-hot encoded) variables, setting $x_{\mathrm{node}}^{[\tau]}=1$ if we visit $\mathrm{node}=[s,d,t,c,p]$ at position $\tau=1, \dots, n_{\mathrm{seams}}$ of the tour, and $x_{\mathrm{node}}^{[\tau]}=0$ otherwise. 
Following the QUBO formulation for the canonical TSP problem \citep{lucas:14}, we can then describe the goal of finding a minimal-time tour with the quadratic Hamiltonian 
\begin{equation}
H_\mathrm{cost} = \sum_{\tau=1}^{n_{\mathrm{seams}}} \sum_{\mathrm{node}} \sum_{\mathrm{node}'} w_{\mathrm{node}}^{\mathrm{node}'} x_{\mathrm{node}}^{[\tau]} x_{\mathrm{node}'}^{[\tau+1]}, \label{eq:H-cost}
\end{equation}
with $w_{\mathrm{node}}^{\mathrm{node}'}$ denoting the cost to go from node to $\mathrm{node}'$. 
Here, the product $x_{\mathrm{node}}^{[\tau]} x_{\mathrm{node}'}^{[\tau+1]}=1$, if and only if node is at position $\tau$ in the cycle and $\mathrm{node}'$ is visited right after at position $\tau+1$. In that case we add the corresponding distance $w_{\mathrm{node}}^{\mathrm{node}'}$ to our objective function which we would like to minimize. 
Overall, we sum all costs of the distances between successive nodes.
Next, we need to enforce the validity of the solution through additional penalty terms, i.e., we need to account for the following constraints: 
First, we should have exactly one node assigned to every time step in the cycle. 
Mathematically, this constraint can be written as 
\begin{equation}
\sum_{s,d,t,c,p} x_{[s,d,t,c,p]}^{[\tau]} = 1, \quad \forall \tau = 1, \dots, n_{\mathrm{seams}}.    
\end{equation}
Second, every seam should be visited once and only once (in some combination of the
remaining features). Note that we do not have to visit every potential node.
This constraint is mathematically captured by 
\begin{equation}
\sum_{\tau}\sum_{d,t,c,p} x_{[s,d,t,c,p]}^{[\tau]} = 1, \quad \forall s=1, \dots, n_{\mathrm{seams}}.    
\end{equation}
As detailed in Ref.~\citep{glover:18} within the QUBO formalism, we capture these constraints through additional penalty terms given by
\begin{eqnarray}
H_\mathrm{time} & = & P \sum_{\tau=1}^{n_{\mathrm{seams}}} \left[ \sum_{\mathrm{node}} x_{\mathrm{node}}^{[\tau]} - 1\right]^2, \\ 
H_\mathrm{complete} & = & P \sum_{s=1}^{n_{\mathrm{seams}}} \left[ \sum_{\tau}\sum_{d,t,c,p} x_{[s,d,t,c,p]}^{[\tau]} - 1\right]^2, 
\label{eq:H-penalty}
\end{eqnarray}
with the penalty parameter $P>0$ enforcing the constraints. Note that the numerical value for $P$ can be optimized in an outer loop. 
Finally, the total Hamiltonian describing the use case then reads
\begin{equation}
H_\mathrm{RM} = H_\mathrm{cost} + H_\mathrm{time} + H_\mathrm{complete}. \label{eq:H-rm-qubo}
\end{equation}
Because the Hamiltonian $H_\mathrm{RM}$ is quadratic in the binary decision variables $\{x_{\mathrm{node}}^{[\tau]}\}$, it falls into the broader class of QUBO problems, which is amenable to quantum-native solution strategies, for example in the form of quantum annealing \citep{hauke:20}, in addition to traditional classical solvers such as simulated annealing or tabu search.

\subsubsection{Resource estimation} 

We complete this analysis with a rough estimate for the number of (logical) qubits $n_\mathrm{qubits}$ required to implement this QUBO formulation for industry-relevant scales. With the number of time-steps for the Hamiltonian cycle given by $n_{\mathrm{seams}}$ we obtain
\begin{equation}
n_\mathrm{qubits} = 2\times n_{\mathrm{seams}}^{2} \times n_\mathrm{tools} \times n_\mathrm{config} \times n_\mathrm{position}. \label{eq:number-qubits}
\end{equation}
Taking numbers from a realistic industry-scale case we use $n_{\mathrm{seams}} \sim 50$, $n_\mathrm{tools} \sim 3$, $n_\mathrm{config} \sim 10$, and $n_{\mathrm{position}} \sim 3$, we then find $n_\mathrm{qubits} \sim 5 \times 10^{5}$.  
Furthermore, given the quadratic overhead for embedding an all-to-all connected graph onto the sparse Chimera architecture \citep{perdomo:17y}, we can estimate the required number of \textit{physical} qubits $N_\mathrm{qubits}$ to be as large as 
\begin{equation}
    N_\mathrm{qubits}\sim 10^{11}.
\end{equation}
This number is much larger than the number of qubits available today and in the foreseeable future, and higher connectivity between the physical qubits will be needed to reduce our requirements for $N_\mathrm{qubits}$. 
Therefore, in our numerical experiments we utilize hybrid (quantum-classical) solvers which heuristically decompose the original QUBO problem into smaller subproblems that are compatible with today's hardware.
These subproblems are then solved individually on a quantum annealing backend, and a global bit string is recovered from the pool of individual, small-scale solutions by stitching together individual bit strings.

Finally, for illustration purposes let us also compare the size of the combinatorial search space any QUBO-based approach is exposed to, as compared to the native search space underlying the random-key formalism. 
Disregarding (for simplicity) all complementary categorical features for now, the size of the search space for the QUBO formalism amounts to $2^{n_{\mathrm{seams}}^2}$ while a native encoding has to search through the space of permutations of size $n_{\mathrm{seams}}!$.
This means that for $n_{\mathrm{seams}} \sim 50$ the latter amounts to $\sim 10^{64}$ while the QUBO search space is many orders of magnitude larger, with $2^{50^2} \sim 10^{752}$ possible solution candidates, thus demonstrating the benefits of an efficient, native encoding for sequencing-type problems as relevant here. 

\section{Numerical experiments and benchmarks}
\label{Numerics}

We now turn to systematic numerical experiments for industry-relevant datasets. 
We first describe the generation of the datasets, and then compare results achieved with random-key algorithms to baseline results using greedy methods. 
Our implementation of BRKGA is based on code originating from Ref.~\citep{AndTosGonRes21a}.
We also provide results achieved with simulated annealing (SA) applied within the QUBO modelling framework (referred to as QUBO-SA).  
To assess the capabilities of today's quantum hardware, we complement these classical approaches with quantum-native solution strategies, including results obtained on quantum annealing hardware within a hybrid quantum-classical algorithm, using \texttt{qbsolv} on Amazon Braket (referred to as QUBO-QBSolv) \citep{braket}. 

\subsection{Datasets} 

\begin{figure*} [t]
\includegraphics[width=1.9 \columnwidth]{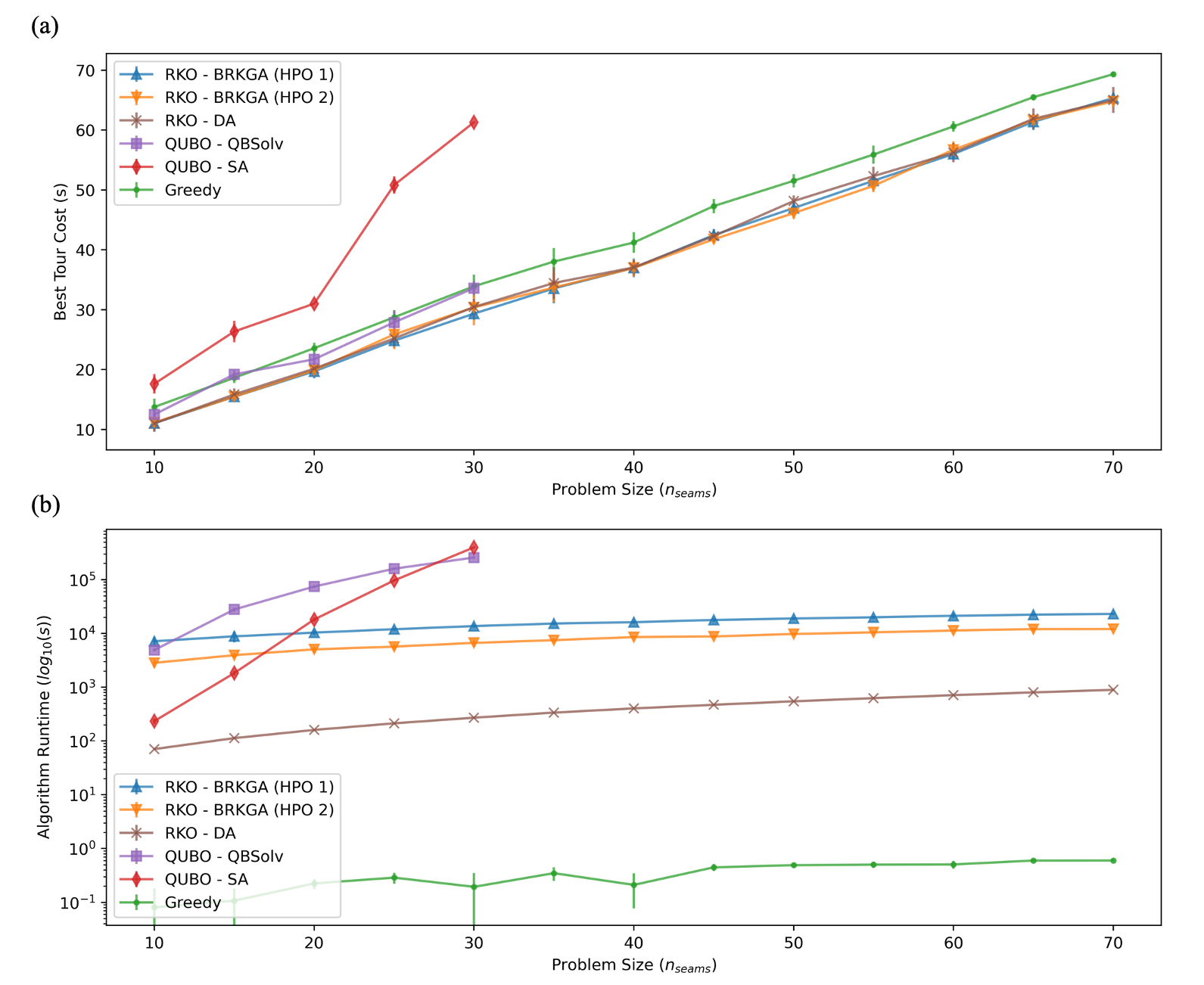}
\caption{
(Color online) 
Numerical results for systems with up to $n_{\mathrm{seams}}=70$ seams. Other dimensions are fixed, with $n_{\mathrm{d}}=2$, $n_{\mathrm{tools}}=3$, $n_{\mathrm{config}}=9$, $n_{\mathrm{position}}=4$.
For $n_{\mathrm{seams}}=20$ the QUBO size as measured by the required number of binary variables amounts to $\sim 10^{5}$, posing limits on the accessible problem size because of memory restrictions.
(a) Solution quality as measured by the total cost (i.e., duration) of the best tour found (in seconds).
All results have been averaged over ten samples per system size, specified by the number of seams.
We compare results achieved with BRKGA for two sets of hyperparamters (as shown in blue upward triangle [HPO 1] and orange downward triangle [HPO 2]) and for RKO-DA (brown x) with one set of hyperparameters, with a simple greedy heuristic serving as baseline (green circles).
For instances with $n_{\mathrm{seams}} \lesssim 20$ we provide results based on the QUBO formalism for both classical simulated annealing (QUBO-SA, red diamonds), as well as a hybrid (quantum-classical) decomposing solver \texttt{qbsolv} (QUBO-QBSolv, purple diamonds).  
(b) Algorithm runtime as a function of the number of seams $n_{\mathrm{seams}}$. 
The greedy baseline algorithm is extremely fast with its runtime showing practically no discernible dependence on the system size in the parameter regime tested here.
Our implementation of BRKGA displays a mild, linear scaling with the absolute numbers depending on hyperparameters such as population size.
Still, the largest problems with $n_{\mathrm{seams}} \sim 70$ seams are solved within a few hours. 
The RKO-DA implementation is found to be about an order of magnitude faster than BRKGA. 
The QUBO-based approaches display comparatively long run times for sufficiently large system sizes. 
Further details are given in the text. 
\label{fig:numerics}
}
\end{figure*}

All datasets were generated by BMW Group using custom logic that is implemented in the Robot Operating System (ROS) framework \citep{ros}.
To be able to calculate the relevant cost values (measured in seconds) to move between nodes and build the cost matrix $W$, the physical robot cell is modelled in ROS.
This includes loading and positioning of the robot model, importing static collision objects and parsing the robotic objectives.
The generation of the dataset is then done in two steps.
First, a reachability analysis is carried out to inspect the different possibilities of applying sealant to a given seam. 
This includes the choice of the nozzle to use, the robot's joint configuration, and the position of the robot on the linear axis.
The latter is discretized to provide a finite number of possible linear axis positions to be searched, with $n_{\mathrm{position}}$ left as a free parameter in the framework. 
Second, the motion planning is carried out to obtain the trajectories for all possibilities to move from one seam to all possibilities of processing any other seam.
Collision avoidance and time parameterization are included in this process.
To this end we adopted the RRT* motion planning algorithm \cite{karaman2011}.
If the motion planning does not succeed, no (weighted) edge is entered in the motion graph.
The runtime to generate the largest dataset presented in this paper is approximately 2.5 days.
These calculations were performed with 70 threads on an Intel Xeon Gold 6154 CPU, running at 3.00 GHz.
The largest dataset considered contains $n_{\mathrm{seams}}=71$ seams, including the home position.
For our scaling and benchmarking analysis (as shown in Fig.~\ref{fig:numerics}) we have implemented a random down-sampling strategy.
By randomly removing seams from the original dataset we have generated synthetic datasets with variable size, ranging from $n_{\mathrm{seams}}=10$ to $n_{\mathrm{seams}}=70$ in increments of five seams, with ten distinct downsamples per system size.
For reference, within the QUBO formalism the smallest instance with $n_{\mathrm{seams}}=10$ already requires approximately $\sim 10^{4}$ binary variables, thus representing a large QUBO problem to be solved.

\begin{table*}
\centering
\caption{
Numerical results for two industry-relevant benchmark datasets, referred to as benchmark L and benchmark XL, respectively.
For reference, with QUBO size we report the approximate number of binary variables [cf. Eq.~\eqref{eq:number-qubits}] to describe this instance within the QUBO formalism.
We report cost values achieved with three different algorithmic strategies (greedy, BRKGA, and RKO-DA). 
The greedy algorithm has been run $10^{4}$ times, and the best (lowest cost) solution is reported in the table.
Best results across algorithms are marked in bold. 
The last two columns specify the absolute and relative improvement of the best solution over the greedy baseline strategy. 
Further details are provided in the main text.
\label{tab:benchmarks}} 
 \begin{tabular}{|c | c c | c c c | c c |} 
 \hline
 \hline
 Dataset & number of seams & QUBO size & Greedy & BRKGA & RKO-DA & absolute $\Delta$ (s) &  relative $\Delta$ (\%) \\ [0.5ex] 
 \hline
 Benchmark L & 52 & $6\times 10^{5}$ & 37.05 & 33.81 & \textbf{33.30} & 3.75 & 10.12\% \\ 
 Benchmark XL & 71 & $1\times 10^{6}$ & 65.99 & 65.09 & \textbf{63.60} & 2.39 & 3.62\% \\ [1ex] 
 \hline
 \hline
 \end{tabular}
\end{table*}

\subsection{Benchmarking results} 

First we provide results for two selected, large-scale, industry-relevant benchmark datasets, referred to as benchmark L and benchmark XL with $n_{\mathrm{seams}}=52$ and $n_{\mathrm{seams}}=71$, respectively.  
While benchmark XL represents the largest available problem instance, benchmark L, while smaller, represents a particularly hard instance because it features many obstructions to the robot's path (resulting in missing entries in the motion graph).   
We compare three different algorithmic strategies on these benchmark instances, including BRKGA and RKO-DA (both based on the random-key formalism, but following different optimization paradigms), as well as a greedy baseline. 
Here we do not provide results for any QUBO-based solution strategies, because these instances are out of reach for QUBO solvers, with approximately $\sim 6\times10^{5}$ and $10^{6}$ binary variables, respectively. Note that numerical results based on the QUBO formalism will be provided below. 
Hyperparameter optimization for the BRKGA and RKO-DA solvers is done as outlined in Ref.~\cite{BerYamCox2013a} using Bayesian optimization techniques. 
The greedy algorithm was run $10^{4}$ times, each time with a different random starting configuration, thereby effectively eliminating any dependence on the random seed. We have observed convergence for the best greedy result after typically $\sim 10^3$ shots, and only the best solutions found are reported. 
Our results are displayed in Tab.~\ref{tab:benchmarks}. 
We find that both BRKGA and RKO-DA outperform the greedy baseline. 
Specifically, BRGKA yields an improvement of 3.24s (8.74\%) on benchmark L and 0.90s (1.36\%) on benchmark XL, while RKO-DA yields an even larger improvement of 3.75s (10.12\%) on benchmark L and 2.39 (3.62\%) on benchmark XL. 
These improvements can directly translate into cost savings, increased production volumes or both.

\subsection{Scaling results} 

To complement our benchmark results, we have performed systematic experiments on datasets with variable size, ranging from $n_{\mathrm{seams}}=10$ to $n_{\mathrm{seams}}=70$ in increments of five seams, with ten samples per system size.
Our results are displayed in Fig.~\ref{fig:numerics}, with every curve referring to one fixed set of hyperparameters (such as population size $p$, mutant percentage $p_{m}$, etc. in the case of BRKGA). 
Further details regarding hyperparameters can be found in the Appendix.
We also report numerical results based on the quantum-native QUBO formalism, using both classical simulated annealing, as well as a hybrid (quantum-classical) decomposing solver \texttt{qbsolv}. 
However, for $n_{\mathrm{seams}}=20$ the QUBO size as measured by the required number of binary variables already amounts to $\sim 10^{5}$, posing limits on the accessible problem size because of memory restrictions.
We compare results achieved with BRKGA for two sets of hyperparamters, and for RKO-DA with one set of hyperparameters, and find that these consistently outperform greedy baseline results, providing about $\sim 10\%$ improvement for real-world systems with $n_{\mathrm{seams}} \sim 50$. Note that BRGKA (HPO 2) was run with ten shots for problem sizes 45, 50, 55, and 70 to reduce variance observed when using fixed hyperparameters. 
Performance of RKO-DA is found to be competitive with BRKGA throughout the range of problem sizes.
The QUBO-SA solver is found to be the least competitive, and the least scalable, unable to solve problems beyond $n_{\mathrm{seams}} \approx 20$. 
Finally, the QUBO-QBSolv approach performs on par with the greedy algorithm, albeit at much longer run times, but is unable to scale beyond $n_{\mathrm{seams}} \approx 30$. 

Next, we report on  the average run times of each solver as a function of problem size; see Fig.~\ref{fig:numerics} (b).
Given its simplicity, the greedy algorithm is found to be the fastest solver, taking under a second ($\sim 0.60$s) to solve the largest problem instances with $n_{\mathrm{seams}}=70$, after loading in the requisite data. 
Our implementation of the random-key optimizer RKO-DA can solve the largest problem instances within $\sim 15$ minutes, and exhibits a benign, linear runtime scaling with problem size. 
Our implementation of BRKGA displays a similar scaling, but typically takes a few hours to complete, with average run times ranging from $\sim 0.78$h (HPO 2, $n_{\mathrm{seams}}=10$) to $\sim 6.40$h (HPO 1, $n_{\mathrm{seams}}=70$). 
The observed difference in run times between BRKGA (HPO 1) and BRKGA (HPO 2) can be largely attributed to the different population sizes (with $p=9918$, and $p=7978$ for HPO 1 and HPO 2, respectively) and patience values (with $\mathrm{patience}=52$, and $\mathrm{patience}=66$ for HPO 1 and HPO 2, respectively).
The observed speed-up from BRKGA to RKO-DA, both based on the random-key formalism, can be attributed to the internal anatomy of the respective algorithm: 
While BRKGA evolves entire populations of (in our case typically $p\sim 10^{4}$) interacting chromosomes, RKO-DA tracks only one single chromosome $\mathcal{X}$ throughout its algorithmic evolution, yielding about an order of magnitude speed-up across the range of problem sizes studied here.
Finally, as compared to the heuristics described above, the QUBO-based approaches display unfavorable run times, either because of orders-of-magnitude longer run times across the range of accessible problem sizes (QUBO-QBSolv) or because of unfavorable runtime scaling (QUBO-SA), further corroborating the advantage of our native solution strategies as compared to quantum-native QUBO formulations.

\begin{figure*} [t]
\includegraphics[width=2.1 \columnwidth]{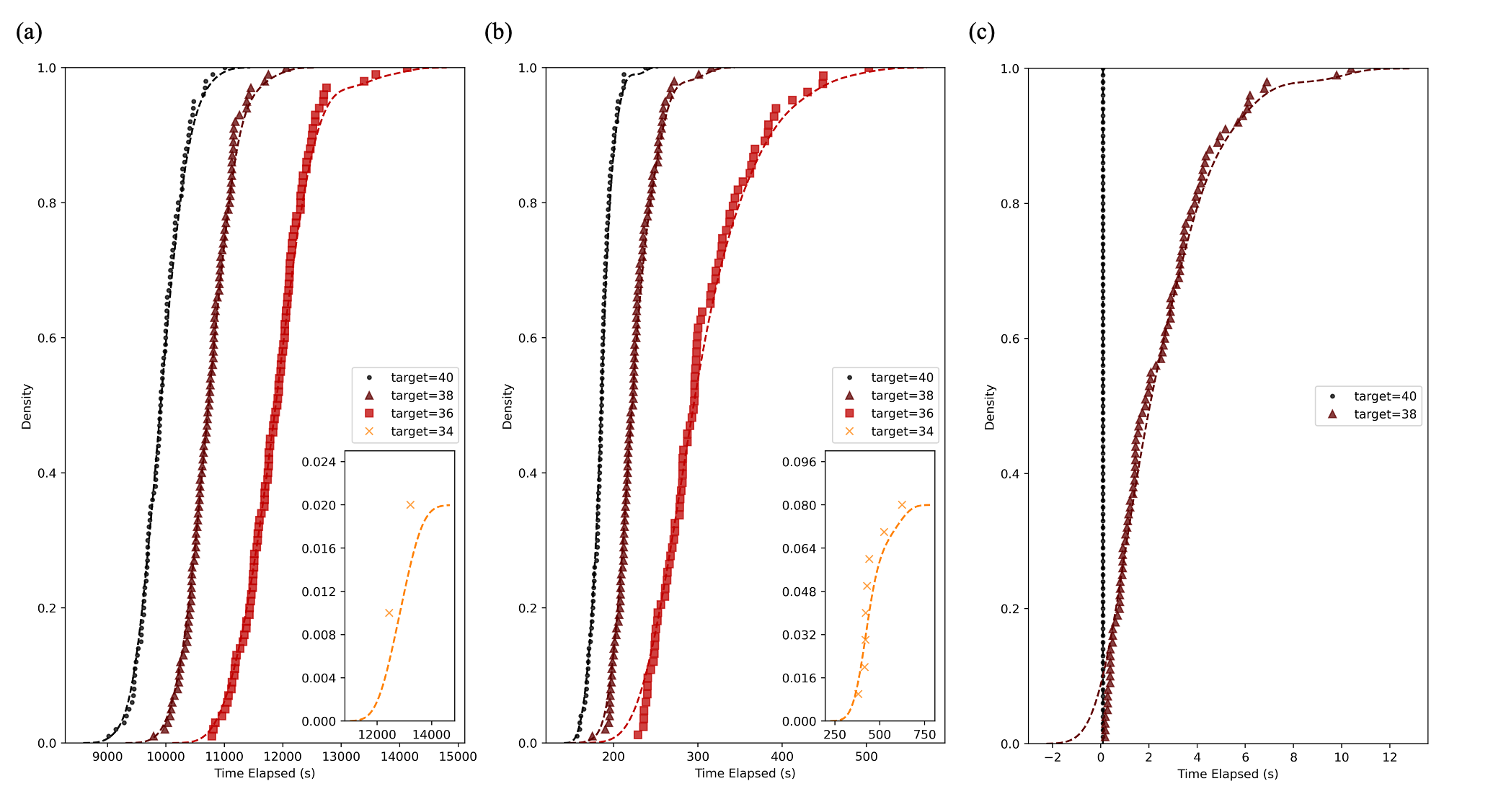}
\caption{
(Color online) Time-to-target (TTT) plots for (a) BRKGA, (b) RKO-DA, and (c) greedy solvers on the benchmark L dataset with $n_{\mathrm{seams}}=52$. 
Target values (measured in seconds) have been set to 40 (poor solution quality; black circles), 38 (Brown triangle), 36 (red square), and 34 (high solution quality; orange x); compare Tab.~\ref{tab:benchmarks}.
Plots for a target of 34 (if available) are placed in the insets for clarity.
We plot the probability to hit a fixed target cost value on the vertical axis, as a function of the algorithm's runtime on the horizontal axis. 
Each algorithm has been run $100$ times to convergence with fixed hyperparameters, varying the seed on each run, and tracking every intermediate step. 
The greedy solver is unable to reach or exceed a target of 37 (the best greedy solution is $37.05$), while 
both BRKGA and RKO-DA reach a target of 34, with probability $\sim 2\%$ for BKRGA, and $\sim 8\%$ for RKO-DA.
Further details are given in the text. 
}
\label{fig:ttt_plots}
\end{figure*}

\subsection{Time to target results} 

We complete our numerical benchmark analysis with time-to-target (TTT) studies, as outlined and detailed (for example) in Ref.~ \citep{resende:03}.
TTT plots have been shown to be useful in the comparison of different algorithms, and have been widely used as a tool for algorithm design and comparison \citep{resende:07}.
As described in Ref.~\citep{resende:07}, on the ordinate axis TTT plots display the probability that an algorithm will find a solution at least as good as a given target value within a given runtime (shown on the abscissa axis), thereby characterizing expected run times of \textit{stochastic} algorithms within empirical (cumulative) probability distributions. 
We extract these by measuring CPU run times needed to find a solution with objective function value at least as good as a fixed target value, for a given problem instance. 
To this end we run a given algorithm $n_{\mathrm{shots}}$ times, with a distinct seed for every run (thus giving independent runs).
Our results are displayed in Fig.~\ref{fig:ttt_plots}, showing TTT plots for BRGKA, RKO-DA, and greedy solvers on benchmark L with $n_{\mathrm{shots}}=100$ and several target values. 
The results show that each algorithm will take longer to hit a given target as this target cost becomes smaller. 
Furthermore, the greedy solver is unable to reach or exceed a target of 37 (the best greedy solution is $37.05$), while 
both BRKGA and RKO-DA reach a target of 34, with probability $\sim 2\%$ for BKRGA, and $\sim 8\%$ for RKO-DA, respectively.
Finally, the expected runtime across solvers, for any given target value, is different by orders of magnitude: greedy is fastest, RKO-DA is next fastest, and BRKGA is slowest. 
For example, for RKO-DA a runtime of approx.~$\sim 300$s is sufficient to find a target solution with cost of $36$ (or better) with approx.~50\% success probability, while BRKGA needs about $1.2\times 10^4$s to achieve the same while the greedy algorithm completely fails to achieve this solution quality.   \section{Conclusion and Outlook}
\label{Conclusion}

In summary, we have shown how to solve robot trajectory planning problems at industry-relevant scales. 
To this end we have developed an end-to-end optimization pipeline which integrates classical random-key algorithms with quantum annealing into a quantum-ready, future-proof solution to the problem.
With the help of a distinct separation of problem-independent and problem-dependent modules our approach achieves an efficient problem representation that provides a native encoding of constraints, while ensuring flexibility in the choice of the underlying optimization algorithm. 
We have provided numerical benchmark results for industry-scale datasets, showing that our approach consistently outperforms greedy baseline results. We complement this analysis with a transparent assessment of the capabilities of today's quantum hardware and resource estimates for the number of qubits required to implement a quantum-native problem formulation for industry-relevant scales. 

Finally, we highlight possible extensions of research going beyond our present work, where we have focused on settings involving a \textit{single} robot, as relevant for situations in which collisions between robots can be avoided through the definition of appropriate bounding boxes.
In future work it will be instructive how to generalize our random-key approach towards \textit{multi-robot} problems. 
To this end, we make use of an apparent analogy to the vehicle routing problem (VRP) as outlined in Sec.~\ref{preliminaries}. 
Specifically, we can associate individual robots with vehicles in the traditional VRP. 
With a straightforward extension of our formalism we can then solve both the allocation and the sequencing problems within one unified framework. 
For example, consider a setup with $n_\mathrm{seams}=6$ and two robots, and a sample random-key vector $\mathcal{X}=(0.25, 0.19, 0.67, 0.98, 0.04, 0.82, \mathbf{0.23}, \mathbf{0.71})$.
Here, the first $n_\mathrm{seams}=6$ keys correspond to seams to be sealed by $v=2$ robots. 
Along the lines of the VRP presented above, this random-key vector translates into a solution where the first robot leaves its home position and visits seams $6, 4, 5, 2$ (in this order), and then return to its base, while the second robot leaves its home position and visits locations $1, 3$ before returning to its base. 
The decoder then computes the corresponding cost value by adding individual cost values, in addition to potential large penalties if the proposed solution incurs a collision between robots. 
Through the feedback mechanism inherent to our approach, the latter will steer the algorithm towards collision-free solutions over the course of the evolution. 

\begin{acknowledgments}

We would like to thank Alexander Opfolter and Cory Thigpen for management of this collaboration between AWS and BMW. The AWS team thanks Shantu Roy for his invaluable guidance and support.

\end{acknowledgments}

\onecolumngrid

\appendix

\section{Example Decoder Design} \label{decoder}

\begin{mdframed}[backgroundcolor=gray!10]
\lstset{language=Python}
\lstset{frame=lines}
\lstset{caption={Core code block of example decoder.}}
\lstset{label={listing:code-block}}
\lstset{basicstyle=\footnotesize}

\begin{lstlisting}[language=Python]
def decode_piecewise(self, chromosome):
    # Split chromosome into (N) dimensions, to be decoded independently
    chr_pieces = np.array_split(chromosome, self.instance.num_dims)

    decoded_pieces = []
    sort_order = None
    for idx, piece in enumerate(chr_pieces):
        if idx == 0: # assume dim0 = abstract node number (e.g. seam number)
            # Sort in ascending order and use the order of indices
            permutation = np.argsort(piece)
            # Track seam order to pair correctly with other dimensions
            sort_order = copy(permutation)
        else:
            # Assume categorical values for all other dimensions
            n_bins = self.instance.dim_sizes[idx]
            step_size = 1. / n_bins
            # Define bin edges
            bins = np.arange(0.0, 1.0+step_size, step=step_size) 
            # Assign values to bins, report bin assignment; offset by 1
            permutation = np.digitize(piece, bins) - 1 
            # Rearrange output according to seam argsort order
            permutation = permutation[sort_order] 

        decoded_pieces.append(permutation)

    # Pair elements in same position across dimensions
    # i.e. [1,2,3], [10,20,30] -> [1,10], [2,20], [3,30]
    decoded = list(zip(*decoded_pieces))
    return decoded

def decode(self, chromosome, rewrite: bool = False) -> float:
    # Decode chromosome into tour, with nodes of N-dimensions
    decoded = self.decode_piecewise(chromosome)

    # Add distance from LAST node to FIRST node
    cost = self.instance.distance(decoded[-1], decoded[0])

    # Cumulative sum of distances for intermediate nodes of tour
    for i in range(len(decoded)-1):
        cost += self.instance.distance(decoded[i], decoded[i + 1])

    return cost
\end{lstlisting}
\end{mdframed}

\section{Hyperparameters for Numerical Experiments} \label{hypers}

\begin{table*}[h]
\centering
 \begin{tabular}{| c | c | c | c | c | c | c | c | c | c |} 
 \hline
 benchmark & elite\_percentage & mutants\_percentage & num\_generations & patience & population\_size & seed & num\_elite\_parents & total\_parents \\ [0.5ex]
 \hline
 L & 0.4465 & 0.0518 & 2000 & 52 & 9918 & 839 & 2 & 3 \\ 
 XL & 0.4894 & 0.2594 & 2000 & 66 & 7978 & 263 & 2 & 3 \\ [1ex] 
 \hline
 \end{tabular}
\caption{
Optimal hyperparameter values (rounded to 4 decimal points) for BRKGA solver on Benchmarks L and XL. 
\label{tab:hypers_brkga}} 
\end{table*}

\begin{table*}[h]
\centering
 \begin{tabular}{| c | c | c | c | c | c | c | c | c | c |} 
 \hline
 benchmark & maxiter & seed & visit & accept & initial\_temp & restart\_temp\_ratio \\ [0.5ex]
 \hline
 L & 5547 & 151 & 1.1321 & -2.3875 & 20314.2789 & 6.3192$e^{-5}$ \\ 
 XL & 27635 & 656 & 1.1741 & -0.4968 & 49061.6875 & 1.1119$e^{-4}$  \\ [1ex] 
 \hline
 \end{tabular}
\caption{
Optimal hyperparameter values (rounded to 4 decimal points) for RKO-DA solver on Benchmarks L and XL.
\label{tab:hypers_da}}
\end{table*}

\twocolumngrid

\FloatBarrier

\bibliography{references}

\end{document}